\documentclass[twocolumn,trackchanges,twocolappendix]{aastex701}

\usepackage{subcaption} 
\usepackage{caption}
\usepackage{multirow}
\usepackage{booktabs}

\shorttitle{X-ray spectral study of PDS~456}
\shortauthors{Fathima et al.}

\begin{document}

\title{Disk-corona Energy Balance Evolves with Flux State in the Quasar PDS~456} 

\author[orcid=0009-0004-2283-7511,sname='Fathima']{Fida Fathima S S}
\affiliation{School of Astrophysics, Presidency University, 86/1 College Street, Kolkata 700073, India}
\email[show]{fidafathima2018@gmail.com}  

\author[orcid=0009-0005-6387-4789,gname=Arijit, sname='Sar']{Arijit Sar} 
\affiliation{School of Astrophysics, Presidency University, 86/1 College Street, Kolkata 700073, India}
\email{arijitsar8@gmail.com}

\author[orcid=0000-0001-5647-3366, gname=Tathagata,sname=Saha]{Tathagata Saha}
\affiliation{Inter-University Center for Astronomy and Astrophysics, Pune 411007, India}
\affiliation{School of Astrophysics, Presidency University, 86/1 College Street, Kolkata 700073, India}
\email{tathagata@camk.edu.pl}

\author[orcid=0000-0001-9899-7686, sname=Chatterjee,gname=Ritaban]{Ritaban Chatterjee}
\affiliation{School of Astrophysics, Presidency University, 86/1 College Street, Kolkata 700073, India}
\email{ritaban.astro@presiuniv.ac.in}

\begin{abstract}

We investigate the long-term X-ray spectral evolution of the highly luminous, radio-quiet quasar PDS~456 using \textit{AstroSat} observations obtained in 2019 and 2021 together with archival \textit{XMM-Newton}, \textit{NuSTAR} and \textit{Suzaku} observations spanning 2001–2024. The \textit{AstroSat} observations provide valuable coverage following the bright flaring state observed in 2018, revealing that the source underwent a relatively quiescent phase in 2019 before re-brightening in 2021. During the \textit{AstroSat} epochs, PDS 456 exhibits a harder-when-brighter spectral trend, whereas the long-term \textit{XMM-Newton}, \textit{NuSTAR} and \textit{Suzaku} observations confirm the overall softer-when-fainter behavior. This behavior is consistent with the presence of a very cool corona ($\sim15$~keV), possibly driven by a strong incident photon field for up-scattering, as suggested by a prominent big blue bump. Broadband spectral modeling of simultaneous \textit{XMM-Newton} and \textit{NuSTAR} observations, complemented by the 2021 \textit{AstroSat} flaring epoch and interpreted within the self-consistent physical framework \textsc{reXcor}, indicates that the disk–corona energy balance evolves with flux. In low-flux states, the spectrum is reflection-dominated, implying efficient illumination of the inner accretion disk by a compact corona. In contrast, high-flux states show an enhanced contribution from the warm Comptonizing region at larger coronal heights. The Fe K$\alpha$ emission line displays a positive correlation with the 2–10~keV continuum flux at low to intermediate flux levels, while deviations at the highest flux states suggest a reduced relative reflection contribution as the warm corona becomes energetically dominant.
\end{abstract}


\section{Introduction}
Active galactic nuclei (AGN), powered by the accretion of matter onto central supermassive black holes (SMBHs), constitute the most luminous persistent sources in the universe \citep{fabian_1999, Padovani_2017}. Their X-ray emission varies on timescales from hours to decades, indicating an origin in a compact region near the SMBH \citep{Mushotzky_1993, Markowitz_2004, Barr_1986}. The geometry and size of this region, known as the hot corona, remain uncertain \citep{Haardt_1991, Haardt_1994}. X-ray spectral studies are therefore essential to constrain the physical conditions, dynamics, and geometry of this central engine.

AGN X-ray spectra generally display a power-law continuum with a high-energy cutoff \citep{Mushotzky_1982, Turner_1989, Zdziarski_2000}, produced by inverse Compton scattering of accretion-disk optical/UV photons by hot coronal electrons ($\sim$100~keV) \citep{Haardt_1993}. A common deviation is the soft X-ray excess below $\sim$2~keV \citep{Arnaud_1985, Singh_1985}, whose origin remains debated \citep{Done_2012}. Proposed explanations include relativistically blurred disk reflection and Comptonization of disk photons in an optically thick warm corona \citep{Crummy_2006, Done_2012, Garcia_2019, Xu_2021, Ghosh_2021, Ghosh_2022}. The latter successfully describes several Seyfert galaxies within the two-corona framework, where a warm, optically thick corona produces the soft excess and a distinct hot corona generates the hard X-ray continuum  \citep{Petrucci_2013, Petrucci_2018, Jin_2017, Middei_2018, Middei_2019, Ursini_2020}. The reflection signatures, including the Fe K$\alpha$ emission line at 6.4~keV \citep{Pounds_1989} and the Compton reflection hump above 10~keV \citep{George_1991} arise from illumination of the accretion disk by coronal X-ray emission. Recent broadband studies indicate that combining warm and hot Comptonization with relativistic reflection can provide a more complete description of AGN X-ray spectra than either process alone. \citep{Matt_2014, Porquet_2018, Porquet_2021, Matzeu_2020}

The radio-quiet quasar PDS~456 is renowned for its exceptional luminosity and dynamic behavior. It is the most luminous quasar in the local Universe ($z~=~0.184$) \citep{Simpson_1999} discovered during the Pico dos Dias (PDS) Survey of bright IRAS sources \citep{Torres_1997}, with a bolometric luminosity of \(\sim10^{47}~\text{erg~$s^{-1}$}\) \citep{Reeves_2000}. This extraordinary luminosity suggests that the source is accreting at a very high rate \citep{Reeves_2009, Reeves_2021, Gravity_2023}. Its black hole mass was long estimated to be $10^9~M_\odot$ \citep{Reeves_2009, Nardini_2015}, using empirical scaling relations from reverberation mapping studies. However, recent measurements have revised this estimate downward to $10^8~M_\odot$ \citep{Gravity_2024, li_2024}.

A blue-shifted Fe K-shell absorption feature detected in the 2001 {\it XMM-Newton} observation provided evidence for an ultra-fast outflow \citep{Reeves_2003}. Subsequent observations revealed that this highly ionized wind exhibits velocities between \(0.25c\) and \(0.3c\) \citep{Reeves_2003, Reeves_2009, Reeves_2014, Reeves_2018b, Reeves_2018a, Reeves_2020, Behar_2010, Gofford_2014, Hagino_2015, Nardini_2015, Matzeu_2016, Matzeu_2017a, Parker_2018a, Boissay-Malaquin_2019}. Long-term monitoring linked the outflow velocity to intrinsic hard X-ray luminosity, supporting a radiatively driven disk wind \citep{Matzeu_2017b} while recent {\it XRISM} observations revealed multiple velocity components, suggesting a clumpy and inhomogeneous outflow \citep{Xrism_2025}. Variability studies by \cite{Gofford_2014} constrained the absorber's location to within $10^2 - 10^3~\rm R_g$ of the black hole.

PDS~456 exhibits rapid X-ray variability, first identified in the 1998 \textit{RXTE} observation, when the 2--10~keV flux doubled within 17~ks, which is unusual for such a highly luminous quasar \citep{Reeves_2000}, constraining the X-ray emitting region to a few gravitational radii \citep{Reeves_2002}. Similar rapid variability was observed in the 2007 and 2013 \textit{Suzaku} observations, with the most remarkable event occurring in March 2013, when the X-ray flux rose nearly threefold within a day \citep{Matzeu_2016, Matzeu_2017a}, while the bright 2018 observations revealed a spectrum lacking strong wind signatures and an unusually cool X-ray corona \citep{Reeves_2021}. 

Long-term variability of the broad Fe K$\alpha$ emission line and its connection to the X-ray continuum have been investigated in a limited number of AGN, including NGC~1566 \citep{LIANG_2022}, NGC~2992 \citep{murphy_2007}, MCG-6-30-15 \citep{ballantyne_2002, Marinucci_2014}, NGC~4051 \citep{ponti_2006, wang_1999}, and IRAS~13224-3809 \citep{fabian_2013, Parker_2017, jiang_2018}. These studies reveal source-dependent Fe K$\alpha$ variability, reflecting differences in the geometry, ionization state, and physical origin of the reflecting material. Building on these findings, we investigate the relationship between Fe K$\alpha$ line flux and the X-ray continuum flux in PDS~456 to determine how the line responds to continuum variability.

We present 2019 and 2021 \textit{AstroSat} observations of PDS~456 in different flux states, following its bright 2018 flare, alongside the high-quality archival \textit{XMM-Newton}, \textit{NuSTAR} and \textit{Suzaku} data. These multi-epoch observations allow us to investigate spectral variability, the inner accretion flow, and coronal properties and their evolution across flux states using physically motivated spectral models. The Galactic column density \footnote{ \url{https://www.swift.ac.uk/analysis/nhtot/ }} is fixed at $3.35 \times 10^{21}$~cm$^{-2}$ \cite{Willingale_2013}, and the redshift at 0.184 \citep{Torres_1997}.

The paper is organized as follows: Section \ref{sec:Data} describes the observations and data reduction; Section \ref{sec:analysis} presents the spectral analysis; Section \ref{sec:discussion} summarizes the main results; and Section \ref{sec:conclusion} presents the conclusions.

\section{Observations and Data Reduction} \label{sec:Data}
\begin{table*}[htbp]
\centering
\begin{tabular}{cccccccccccc}
\hline
  Time   &  Obs. Id. & Telescope & Abbreviation & Obs. mode & GTI Exp. $^{a}$ & Counts$^{b}$ & \\
  & & & & &(ks) &\\
  \hline
  2001-02-26  & 0041160101 & {\it XMM-Newton} & XMM1 & --   & 36.21 & $9.48 \times10^4$&\\
  2007-09-12   & 0501580101 & {\it XMM-Newton} & XMM2 & LW & 69.3 & $2.40\times10^5$ &\\
  2011-03-16  & 705041010  & {\it Suzaku} & Sz1 &  --  & 125.5 
     & $2.30 \times 10^4$ \\
  2013-08-27 & 60002032002 & {\it NuSTAR} & Nu1 & -- & 43.8 & $1.18\times10^4$ & \\
  2013-09-06 & 0721010301 & {\it XMM-Newton} & XMM3 & -- & 92.36 &$2.04 \times 10^5$ &\\
  2013-09-06 & 60002032004 & {\it NuSTAR} & Nu2 & -- & 42.9 &$2.98 \times 10^3$ &\\
  2013-09-15 & 60002032006 & {\it NuSTAR} & Nu3 & -- & 44.0 & $6.98 \times 10^3$ &\\
  2013-09-20   & 60002032008 & {\it NuSTAR} & Nu4 & -- & 58.5 & $8.94\times10^3$ &\\
  2014-02-26   & 60002032010 & {\it NuSTAR} & Nu5 & -- & 109.3 & $9.8 \times 10^3$ &\\
  2015-07-21 & 90101008002 & {\it NuSTAR} & Nu6 & -- & 40.8 & $1.1\times10^4$ & \\
  2015-07-24 & 90101008004 & {\it NuSTAR} & Nu7 & -- & 77.2& $5.8 \times 10^3$ & \\
  2017-03-23   & 60201020002 & {\it NuSTAR} & Nu8 &-- & 158.0 & $2.2\times10^4$ & \\
  2018-09-20   & 0830390101  & {\it XMM-Newton}  & XMM4  & -- &  86.0  & $2.54\times 10^5$ &\\
  2018-09-20   & 60402007002 & {\it NuSTAR} & Nu9 & -- & 81.9 & $3.4 \times 10^4$ &\\ 
  2019-09-24   & 60402007004 & {\it NuSTAR} & Nu10 & -- & 78.3 & $7.2 \times 10^3$ & \\
  2019-04-11 &  A05\_046T01\_9000002834 & {\it Astrosat} & As1 & -- & 217.4 & $5.1 \times 10^4$ &\\
  2021-03-26 & A10\_007T01\_9000004276 & {\it Astrosat} & As2 & -- & 67.1 & $2.3\times10^4$ &\\
  2021-04-30 & A10\_007T01\_9000004350 & {\it Astrosat} & As3 & -- & 120.0 & $4.1\times10^4$ &\\
  2024-03-10 & 60901011002 & {\it NuSTAR} & Nu11& -- & 159.2 & $3.3\times10^4$ & \\
  
  \hline
\end{tabular}
    \caption{List of observations .\\
    (a)- For \textit{XMM-Newton} GTI for PN instrument, (b)- For \textit{XMM-Newton} counts for PN instrument only, for \textit{NuSTAR} sum of both the FPM modules, for \textit{Astrosat} only the SXT counts.\\
    The two consecutive AstroSat observations from April 5–10 and April 11–18, 2019, separated by a one-day gap, were treated as a single epoch.}
    \label{tab:Table1}
\end{table*}

\subsection{NuSTAR}
PDS~456 was observed 11 times with {\it NuSTAR} \citep{harrison_2013} over a period of about 11 years, using two co-aligned X-ray telescopes corresponding to the Focal Plane Modules A (FPMA) and B (FPMB). For all observations, the exposure times were approximately equal. Cleaned event files were extracted using the \textsc{NUPIPELINE} task that adopted standard filtering criteria and used the latest calibration files available in the {\it NuSTAR} calibration database v20240311a.

Source and background spectra were extracted from circular regions of radius 75$^{\prime\prime}$, with the background region positioned a few arcminutes away on the same detector chip. Spectra were grouped to a minimum of 30 counts per bin unless stated otherwise. 

\subsection{Suzaku}
{{\it Suzaku} (\citealt{Mitsuda_2007_PASJ_Suzaku}) observed PDS 456 multiple times between 2007 and 2012. For long-term spectral variability analysis, we use one Suzaku observation from 2011, as data from other observatories cover the remaining epochs. We reduced and analyzed the X-ray Imaging Spectrometers (XIS; \citealt{Koyama_2007_PASJ_XIS}), sensitive to 0.3--10 keV, excluding the Hard X-ray Detector (HXD) due to the lack of a significant detection.}

{The XIS comprises three front-illuminated (FI) CCDs (XIS0, XIS2, and XIS3) and one back-illuminated (BI) CCD (XIS1). We downloaded the cleaned XIS event files from the HEASARC {\it Suzaku} archive and followed the standard {\it Suzaku} data-reduction procedures to generate the spectral products. Events recorded in the 3$\times$3 and 5$\times$5 editing modes were combined and screened using the recommended selection criteria. Source and background spectra were extracted from circular regions of radius 1.5$^{\prime}$, with the background taken from an offset region. Spectral products for XIS0 and XIS3 were extracted from the cleaned event files using \texttt{xselect}. Response matrix files (RMFs) and ancillary response files (ARFs) were generated using \texttt{xisrmfgen} and \texttt{xissimarfgen}, respectively. After verifying the consistency of the FI spectra, XIS0 and XIS3 spectra and their corresponding responses were combined into a single XIS-FI spectrum for the observation.}

\subsection{XMM-Newton}

{\it XMM-Newton} \citep{jansen_2001} has observed the source multiple times in the past decade. For this work, we used the 2013 observation to calibrate our model before applying it to the {\it AstroSat} data. For spectral analysis, we used data from the EPIC-pn instrument. During the observation, the EPIC-pn was operated in the Large window (LW) mode. 
We processed the {\it XMM-Newton} EPIC-pn data using XMM-Newton Science Analysis Software version 21.0.0 \citep{gabriel_2004}, following the standard procedure. First, we used the task \texttt{epproc} to create the linearized photon event list. We then filtered the data for high-energy particle events using the \texttt{tabgtigen} task with RATE $\leq$ 0.5 ct~${\rm s}^{-1}$.
We adopted the recommended flag selection of the macros {\tt XMMEA\_EP}. Source and background spectra were extracted from circular regions of radii 40$^{\prime\prime}$ and 120$^{\prime\prime}$ respectively, with the background region placed on the same chip in a source-free area. The absence of significant pile-up was confirmed using \texttt{epatplot}. Finally, we extracted the spectra, response, and ancillary response files from the data using the \texttt{evselect}, \texttt{rmfgen}, and \texttt{arfgen} tasks, respectively. 
However, we extracted spectral files resolved by pattern 0 and pattern 1--4 photon events, as the effective area is known to differ, especially in the soft bands, depending on pattern. The spectra were then grouped to contain at least 25 counts per spectral bin using the \texttt{specgroup} task. 

\begin{figure}[htbp]
    \centering
    \includegraphics[width=1.0\linewidth]{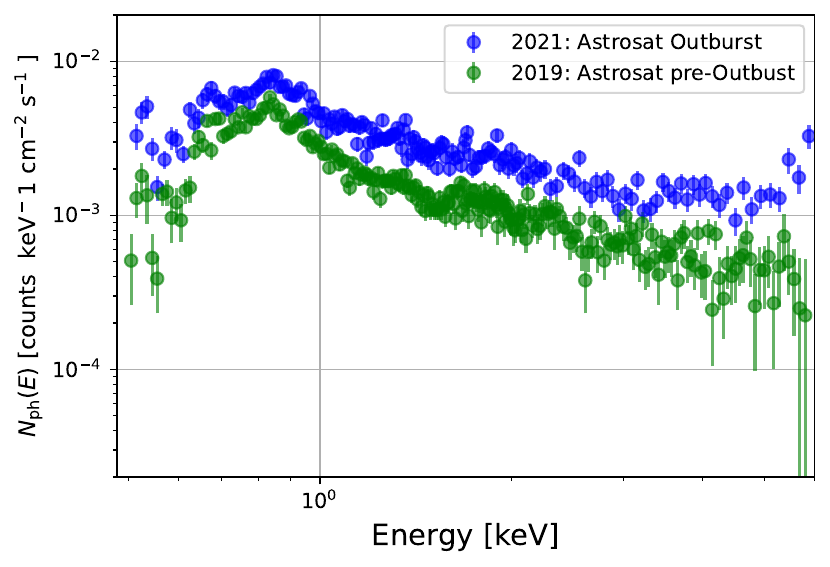}
    \caption{Unfolded AstroSat spectra from the 2019 pre-outburst and 2021 outburst epochs, shown using the $M_{\rm 1,S}$ model.}
    \label{fig:specral_transition}
\end{figure}

\subsection{Astrosat}
{\it AstroSat} \citep{singh_2014} observed PDS~456 multiple times between 2016 and 2021: once in 2016, four times in 2017, twice in 2019, and twice in 2021. We use the 2019 and 2021 Soft X-ray Telescope (SXT) observations aboard \textit{AstroSat}, as the source was not significantly detected in the earlier observations. The two consecutive 2019 observations (April 5–10 and 11–18), separated by a one-day gap, are therefore treated as a single epoch. The 2021 observations were obtained on March 26 and April 30.

The \textit{AstroSat} SXT operated in the photon counting (PC) mode over 0.3-7~keV. Here, we analyzed the 0.7-7~keV band, excluding lower energies due to instrumental features \citep{Das-Chatterjee_2023, Shah_2021}. Level-1 data were retrieved from the {\it AstroSat} archive\footnote{AstroSat data were obtained from the ISSDC AstroSat archive: \url{https://astrobrowse.issdc.gov.in/astro_archive/archive/Home.jsp}.} and processed using the \texttt{sxtpipeline} task of the SXT data processing software AS1SXTLevel2 v1.5 to generate orbit-wise Level-2 cleaned event files \citep{Singh_2016, Singh_2017}, which were merged using the Julia package SXTMergerTool\footnote{SXT Level-2 event files were merged using the Julia-based \texttt{SXTMerger.jl} tool available at \url{https://github.com/gulabd/SXTMerger.jl}.}. The 0.7-7~keV energy range was selected using the \texttt{pha\_cutoff} filter in XSELECT. Source spectra was extracted from a 15$^{\prime\prime}$ radius circular region centered on PDS~456, while background spectra were obtained from the standard blank sky spectrum provided by the SXT Payload Operation Center (POC). The ancillary response file (ARF) and response matrix file (RMF) were obtained from the SXT POC. The extracted spectra were then grouped to a minimum of 25 counts per spectral bin using the \texttt{grppha} task. 
 
Table~\ref{tab:Table1} summarizes the {\it XMM-Newton}, {\it NuSTAR}, {\it Suzaku} and {\it AstroSat} observations used in this work, including observation details, post-filter good time intervals, and net counts. Observation abbreviations are used throughout the spectral analysis, while the Results section identifies epochs by year and flux state to avoid confusion among the numerous datasets.

\section{X-ray Spectral Analysis} 
\label{sec:analysis}

As an initial step, each spectrum was independently fitted using Xspec \citep{Arnaud_1996} with the phenomenological model \textsc{tbabs * zpowerlaw}, where \textsc{tbabs} accounts for galactic photoelectric absorption along the line of sight \citep{Wilms_2000}and \textsc{zpowerlaw} represents the primary coronal X-ray continuum. Significant residuals in the soft (0.3--2~keV) and hard ($>5$~keV) energy bands indicate additional spectral components, including the soft excess and reflection features. To characterize these components and investigate their physical implications, we performed three complementary spectral analyses, each addressing a distinct scientific objective (Table~\ref{QUick_look}). (i) continuum modeling of individual observations using $M_{\rm {1,S}}$ and $M_{\rm {1,H}}$; to characterize the soft- and hard-band spectral continuum; (ii) broadband modeling using $M_{\rm Rex}$ to investigate the interplay between warm Comptonization and relativistic disk reflection within a self-consistent framework; and (iii) phenomenological Fe K$\alpha$ modeling with $M_{\rm Fe}$ to measure the line flux and examine its correlation with the 2–10 keV continuum flux.
\begin{table}[htbp]
\centering
\begin{tabular}{ccccc}
\hline
 Year & Obs. ID & Spectral modelling \\
\hline
2001-02-26   & 0041160101 & $M_{\rm {1,S}}$ $^{a}$; $M_{\rm Fe}$ $^{c}$\\
2007-09-12   & 0501580101 & $M_{\rm {1,S}}$ $^{a}$; $M_{\rm Fe}$ $^{c}$\\
2011-03-16   & 705041010 & $M_{\rm {1,S}}$ $^{a}$; $M_{\rm Fe}$ $^{c}$\\
2013-08-27 & 60002032002 &  $M_{\rm {1,H}}$ $^{b}$; $M_{\rm Fe}$ \\
2013-09-06 & 60002032004 &  $M_{\rm {1,H}}$; $M_{\rm Fe}$; $M_{\rm Rex}$$^{d}$ \\
2013-09-15 & 60002032006 &  $M_{\rm {1,H}}$; $M_{\rm Fe}$\\
2013-09-20   & 60002032008 &   $M_{\rm {1,H}}$; $M_{\rm Fe}$\\
2014-02-26   & 60002032010 &   $M_{\rm {1,H}}$; $M_{\rm Fe}$\\
2015-07-21  & 90101008002 &  $M_{\rm {1,H}}$; $M_{\rm Fe}$\\
2015-07-24 & 90101008004 &  $M_{\rm {1,H}}$; $M_{\rm Fe}$\\
2017-03-23  & 60201020002 &  $M_{\rm {1,H}}$; $M_{\rm Fe}$\\
 2018-09-20   & 60402007002 &  $M_{\rm {1,H}}$; $M_{\rm Fe}$; $M_{\rm Rex}$\\
 2019-09-24   & 60402007004 &   $M_{\rm {1,H}}$; $M_{\rm Fe}$\\
 2019-04-11 &  A05\_046T01\_9000002834 & $M_{\rm {1,S}}$; $M_{\rm Fe}$ \\
2021-03-26  & A10\_007T01\_9000004276 &  $M_{\rm {1,S}}$; $M_{\rm Fe}$\\
2021-04-30 & A10\_007T01\_9000004350 & $M_{\rm {1,S}}$; $M_{\rm Fe}$;  $M_{\rm Rex}$ \\
 2024-03-10 & 60901011002 &  $M_{\rm {1,H}}$; $M_{\rm Fe}$ \\
\hline
\end{tabular}
\caption{Quick-look summary of the spectral modeling applied to each observation.
For all epochs, the source spectra were first fitted using baseline X-ray spectral models to characterize the continuum emission. The Fe K$\alpha$ line flux was then estimated by adding a Gaussian component to the continuum model. Broadband physical modeling was performed only for epochs with simultaneous observations from \textit{XMM-Newton} and \textit{NuSTAR}, as well as for the high-flux state observed with \textit{AstroSat} in 2021.\\
(a) $M_{1,S}= $\textsc{TBabs*zTBabs*(compTT + zpowerlw + uxclumpy\_reflect)} \\
(b) $M_{1,H}= $\textsc{TBabs*(zpowerlw + uxclumpy\_reflect)} \\
(c) $M_{\rm Fe}$= \textsc{tbabs(zpowerlw + zgauss)} \\
(d) $M_{\rm Rex}$= \textsc{Constant*tbabs(Cutoffpowerlaw+reXcor)} for combined \textit{XMM-Newton} and \textit{NuSTAR}, and \textsc{tbabs(zpowerlaw+reXcor)} for 2021 \textit{AstroSat} observation.}
\label{QUick_look}
\end{table}

\subsection{X-ray continuum: the baseline model}
We analyze {\it XMM-Newton}, {\it Suzaku} and {\it Astrosat} spectra in the 0.3--10~keV band using $M_{\rm {1,S}}=$\textsc{TBabs*zTBabs*(compTT + zpowerlw + uxclumpy\_reflect\footnote{XSPEC table spectra of uxclumpy model is available at \url{https://zenodo.org/records/1169181}})}. This model accounts for the soft excess, primary coronal continuum, additional line-of-sight absorption, and possible distant reflection. The \textsc{uxclumpy} \citep{Buchner_2019} describes Compton-scattered emission from a clumpy torus, providing a physically motivated alternative to the slab and toroidal geometries of \textsc{pexmon} \citep{Nandra_2007} and \textsc{MYTorus} \citep{Murphy-Yaqoob_2009}, respectively. We adopt the {\tt uxclumpy} to account for the clumpy and complex structure of parsec-scale circumnuclear reprocessing material \citep{Markowitz_2014, Matzeu_2017a}. For {\it NuSTAR}, we use $M_{\rm {1,H}}=$\textsc{TBabs*(zpowerlw + uxclumpy\_reflect)} over 4-40~keV to constrain the hot-coronal photon index and measure the 4–10~keV flux. We adopt the 4–10~keV band as it is the only energy range common to {\it XMM-Newton}, {\it Suzaku}, {\it Astrosat}, and {\it NuSTAR}, allowing direct comparison of the measured fluxes. 

However, we converted the flux in the 4--10~keV band to 2--10~keV flux phenomenologically since most bolometric corrections are applied to the 2--10~keV band. To convert F(4--10~keV) to F(2--10~keV), we determine the factor by which each F(4--10~keV) flux point varies with respect to XMM2. Under the assumption that the F(2-10~keV) flux points exhibit the same ratio as that of F(4--10~keV), we then multiply this ratio with the F(2--10~keV) flux of XMM2 to get the 2-10~keV flux of each data point. We note a caveat that this flux conversion method does not take into account the variation of the spectrum introduced by a varying soft excess near 2~keV. We do not expect major discrepancies to be introduced by this method since a major contribution of the F(2--10~keV) comes from F(4--10~keV) photons. The Eddington ratio is calculated as $\lambda_{\rm Edd}=L_{\rm bol}/L_{\rm Edd}$, adopting a black hole mass of $10^{8.6} M_{\odot}$ \citep{Gravity_2023}. The bolometric luminosity is estimated using the luminosity-dependent bolometric correction, $L_{\rm bol}=K_{\rm bol}L_{2-10,\rm keV}$ \citep{Lusso_2010, duras2020}, which provides a standardized empirical relation based on a large sample of datasets. Uncertainties in $K_{bol}$, arising from the relation coefficients and the 2-10~keV luminosity, are propagated to estimate the uncertainties in the $L_{bol}$ and $\log\lambda_{\rm Edd}$. The resulting uncertainties are small, with X-ray flux uncertainties of $\lesssim 15\%$ in most epochs and contributions from $K_{\rm bol}$ of approximately $1$--$4\%$. These uncertainties are statistical and do not account for additional systematic uncertainties.

\begin{figure*}[htbp]
    \centering
    \includegraphics[scale=0.52]{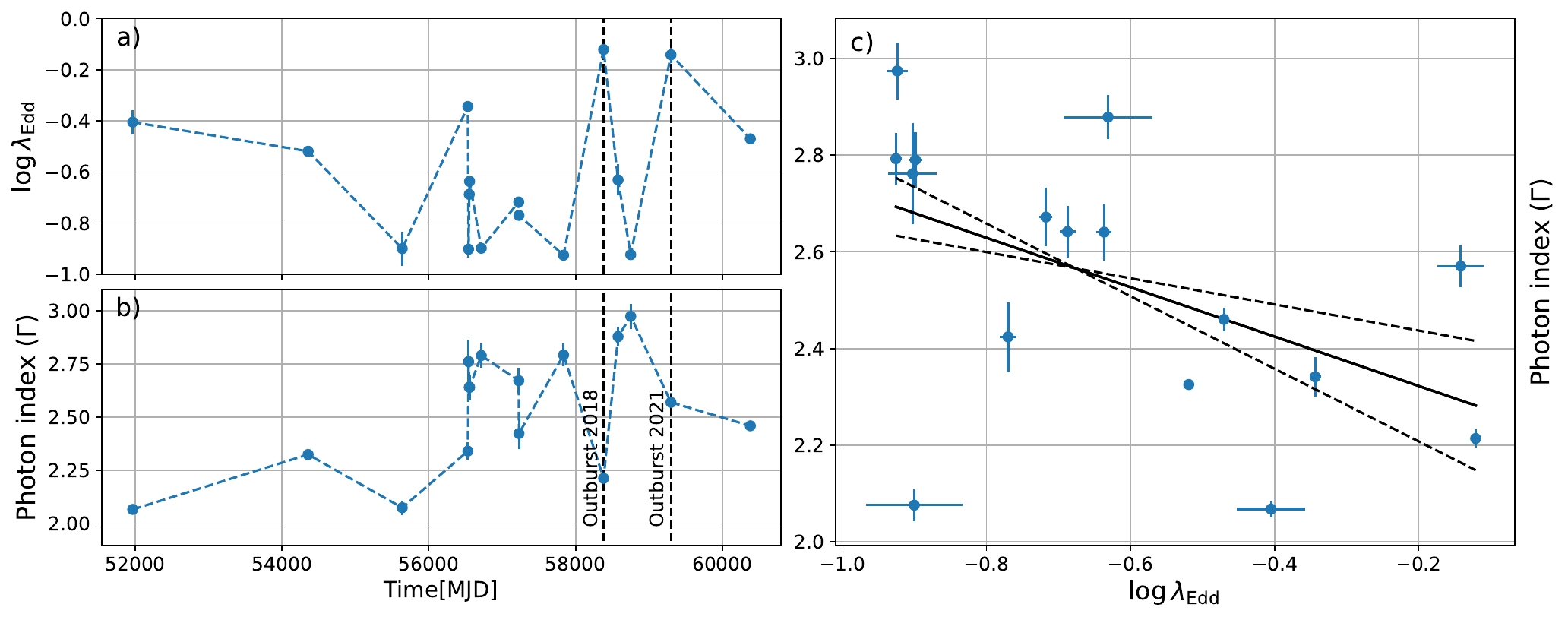}
    \caption{
    (a) Evolution of Eddington ratio with time. A drastic increase in the Eddington ratio is seen in 2018 (outburst reported in \citet{Reeves_2021})and 2021, indicating an outburst 
    (b) evolution of hot-corona photon index with time 
    (c) the evolution of hot-corona photon index with respect to Eddington ratio. These plots indicate that the hot-corona photon index tends to be steeper when the source becomes fainter.}
    \label{fig:residuals}
\end{figure*}

 The derived spectral parameters and fluxes for all observations are summarized in Table \ref{tab:summary_parm}. The unfolded spectra corresponding to the baseline spectral fits are presented in Figure ~\ref{fig:specral_transition} illustrating the spectral differences between the 2019 quiescent (pre-outburst) state and the 2021 April flaring state. To examine the spectral variability across different epochs, we investigate the temporal evolution of the photon index and the Eddington ratio, along with their correlation, as shown in Figure \ref{fig:residuals}.
\begin{table*}[htbp]
\centering
 \begin{tabular}{ccccccc}
 \hline

 Time    & 4--10~keV flux  & 2--10~keV flux  & $\log \lambda_{\rm Edd}$  & $\Gamma$  \\
(MJD)   & ($10^{-12}$~erg~cm$^{-2}$~s$^{-1}$) & ($10^{-12}$~erg~cm$^{-2}$~s$^{-1}$) \\
 \hline
  51965 &  $4.12 \pm 0.7$ &   $7.29\pm0.8$  &  $-0.40\pm0.05$ &  $2.07 \pm 0.02$   \\
  54355.1 &    3.13$\pm$0.03  & 5.84$\pm$0.05  & -0.520$\pm$0.004 &2.33$\pm$0.01 \\
  55637 &  $1.6 \pm 0.4$ &  $2.8 \pm 0.4$  &  $-0.90\pm0.07$  &  $2.08\pm 0.03$\\
  56531.2 &    4.31$\pm$0.08  & 8.1$\pm$0.2     & -0.34$\pm$0.01  &2.34$\pm$0.04\\
  56541.1 &    1.51$\pm$0.12  & 2.8$\pm$0.2     & -0.90$\pm$0.03  &2.8$\pm$0.1\\
  56550.7 &    2.29$\pm$0.06  & 4.3$\pm$0.1    &  -0.69$\pm$0.01  &2.64$\pm$0.05\\
  56555.1 &    2.52$\pm$0.06  & 4.7$\pm$0.1    & -0.64$\pm$0.01  &2.64$\pm$0.06\\
  56714.3 &    1.53$\pm$0.03  & 2.85$\pm$0.06     & -0.90$\pm$0.01  &2.79$\pm$0.06\\
  57224.5 & 2.16$\pm$0.04  & 4.03$\pm$0.08    & -0.72$\pm$0.01  & 2.67$\pm$0.06\\
  57227.5 &    1.96$\pm$0.05  & 3.65$\pm$0.09  & -0.77$\pm$0.01  &2.42$\pm$0.07\\
  57835.2 &    1.45$\pm$0.03  & 2.70$\pm$0.05    & -0.93$\pm$0.01  &2.79$\pm$0.05\\
  58381.1 &    6.42$\pm$0.06  & 12.0$\pm$0.1        & -0.120$\pm$0.004  &2.21$\pm$0.02\\
  58578.0 &    2.54$\pm$0.36  &  4.7$\pm$0.7      & -0.63$\pm$0.06  &2.97$\pm$0.09\\
  58750.4 &    1.45$\pm$0.05  & 2.71$\pm$0.09    & -0.92$\pm$0.01  &2.88$\pm$0.06\\
  59299.0 &    6.19$\pm$0.46  & 11.6$\pm$0.9   & -0.14$\pm$0.03  &2.57$\pm$0.04\\
  60379.9 &    3.43$\pm$0.04  & 6.40$\pm$0.07   & -0.47$\pm$0.01  &2.46$\pm$0.02\\
 \hline
 \end{tabular}
   \caption{Summary of values of flux, Eddington ratio, and photon indices resulting from the fitting of $M_{1,\mathrm{S}}$ and $M_{1,\mathrm{H}}$ (Sec.~3.1). The bolometric correction factor is obtained from \citet{duras2020}.}
   \label{tab:summary_parm}
\end{table*}

\subsection{Spectral fits to the XMM-Newton and NuSTAR data: Connection between Soft-excess and reflection}

Although the baseline model incorporating \textsc{uxclumpy\_reflect} provides a good fit, it does not directly test the interplay between warm Comptonization and relativistic disk reflection as possible contributors to the prominent soft excess in PDS~456. The origin of the soft excess remains uncertain and may arise either independently of, or in connection with, the reflection component, with blurred reflection and warm Comptonization proposed as plausible explanations \citep{Crummy_2006, Boissay_2016}. We therefore employ the physically motivated model \textsc{reXcor} \footnote{\textsc{reXcor} model is freely accessible for spectral fitting with Xspec at \url{https://github.com/HEASARC/xspec_localmodels/tree/master/reXcor}. We used its updated $\Gamma=1.5$--$2.5$ grids available in the repository.} \citep{Xiang_2022} to investigate these processes within a self-consistent framework. By combining warm Comptonization with relativistically blurred ionized reflection, this model allows us to examine their contributions to the broadband X-ray spectrum and their evolution across different flux states.

The \textsc{reXcor} model self-consistently distributes accretion energy among the warm corona, the lamp-post X-ray source at a height $\it{h}$ above the black hole, and the accretion disk, computing the resulting emission while accounting for relativistic effects such as light-bending and gravitational blurring. The predicted soft excess depends on the energy distribution between the warm corona and the lamp-post source. The model provides eight tabulated grids, with fixed combinations of Eddington ratio ($\lambda$), black hole spin (a), and lamp-post height (h). We adopt two grids with $\lambda$= 0.1, the highest available tabulated value and consistent with the high accretion rate of PDS~456 \citep{Gravity_2023}, and a= 0.99, motivated by the high spin measurement of \citep{Chiang_2017}. With these parameters fixed, we compare grids with lamp-post heights of h = 5~$R_{\rm g}$ and 20~$R_{\rm g}$ to investigate the effect of coronal geometry on the broadband spectrum. 

\begin{figure*}[htbp]
    \centering
    \includegraphics[width=1.0\linewidth]{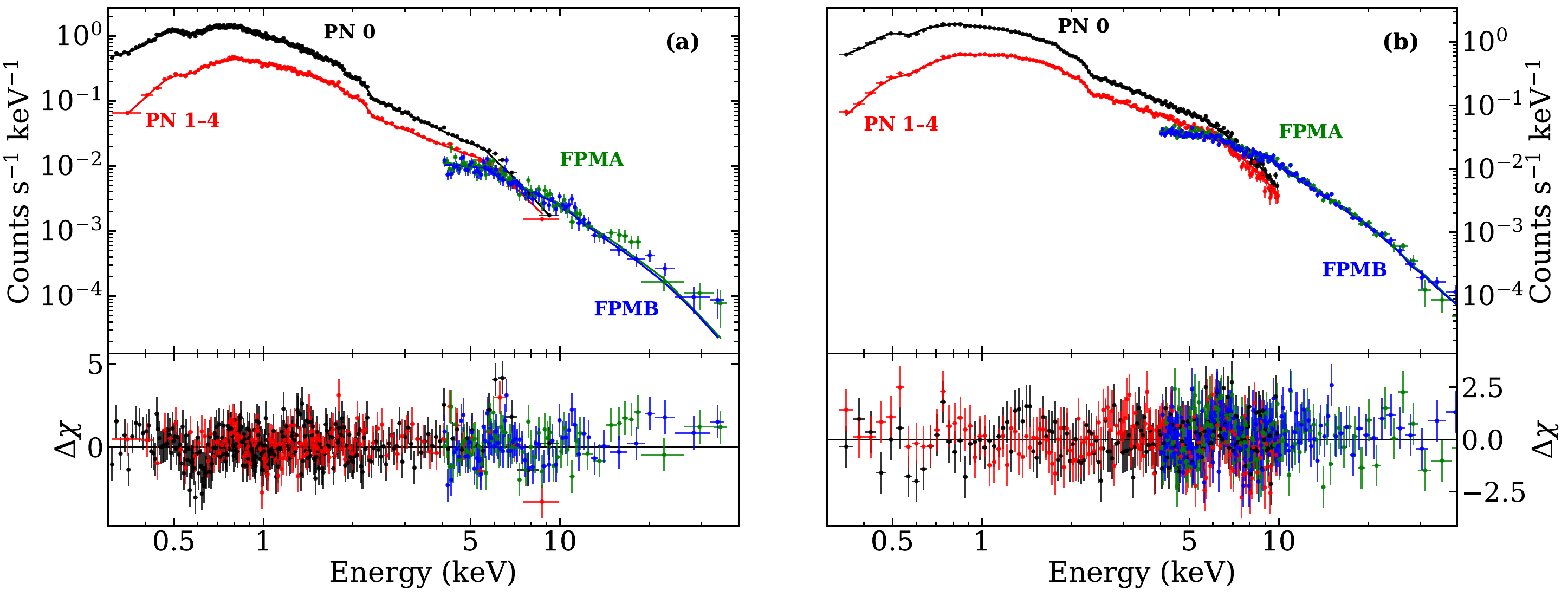}
  \caption{ Broadband \textit{XMM-Newton} + \textit{NuSTAR} X-ray spectra of PDS~456 fitted with the best-fitting\\ \textsc{constant*tbabs(cutoffpl+reXcor} (l=0.1,a=0.99,h=20)) model for two representative flux states. Panel (a) shows the 2013 low-flux observation, while panel (b) shows the 2018 high-flux observation. The lower panels show the (data-model)/error corresponding to the best-fitting model. The PN0, PN1--4, FPMA, and FPMB spectra are shown in black, red, green, and blue, respectively.}
  \label{fig: XMM-Nustar_plots}
\end{figure*}

We applied the selected \textsc{reXcor} grids by jointly fitting two simultaneous \textit{XMM-Newton} and \textit{NuSTAR} observations, corresponding to the 2013 low-flux and 2018 high-flux states. These epochs were selected to compare the disk–corona energy balance across distinct flux states using contemporaneous broadband coverage (0.3–40 keV), which is required to self-consistently constrain the warm-coronal and relativistic reflection components. The observations are XMM3+Nu2 (2013), and XMM4+Nu9 (2018). Both epochs were fitted using \textsc{$M_{\rm Rex}$=Constant*tbabs(Cutoffpowerlaw+reXc\\or)}. The representative best-fit spectra and residuals for the h=20~$R_{\rm g}$ grid are shown in Figure~\ref{fig: XMM-Nustar_plots}, while the fit statistics and best-fit parameters for both grids are summarized in Table~\ref{tab:XMM+NuSTAR_parametrs}.

The normalization constant was fixed at unity for {\it XMM–Newton} and allowed to vary for {\it NuSTAR}. The resulting constant was close to unity, indicating good cross-calibration between the instruments. Best-fit parameters and their associated uncertainties (90\% credible intervals) were estimated using the Markov Chain Monte Carlo (MCMC) method, implemented through the xspec$\_$emcee script\footnote{The MCMC analysis was performed using the \texttt{xspec\_emcee} script, available at \url{https://github.com/jeremysanders/xspec_emcee}.}, which uses the Python-based package \texttt{emcee} \citep{Foreman-Mackey_2013}. The minimum $\chi^2$ value is also reported to indicate the goodness of fit.  

The high-flux state exhibits a relatively low high-energy cutoff of $\sim$50~keV, consistent with \cite{Reeves_2021}. The steeper photon index of $\Gamma$ $\sim$2.3 in the low-flux state decreases to $\Gamma$ $\sim$1.9 in the high-flux state, indicating spectral hardening. At $\it{h}$= 5~$R_{\rm g}$, the optical depth is $\sim11$ in both flux states, and for $\it{h}$= 20~$R_{\rm g}$, it is constrained to an upper limit of 13.5 in the low-flux state, and a lower limit of $\sim19$ in the high-flux state. All parameter limits quoted here correspond to the 95\% credible interval.

The \textsc{reXcor} modeling reveals differences in accretion power distribution between the lamp-post source and the warm corona across epochs and coronal heights. In 2013, the lamp-post heating fraction ($f_{\rm x}$) increases with coronal height, while the warm coronal contribution decreases. In contrast, the warm coronal heating fraction ($h_{\rm f}$) increases significantly at larger coronal heights during the high-flux state of 2018. 

\begin{table*}[htbp]
\centering
\label{tab:rexcor_results}
\begin{tabular}{lcccc}
\hline
Parameter 
& \multicolumn{2}{c}{2013} 
& \multicolumn{2}{c}{2018} \\
\cline{2-5}
\addlinespace
& h = 5~$R_{\rm g}$ & h = 20~$R_{\rm g}$ & h = 5~$R_{\rm g}$ & h = 20~$R_{\rm g}$\\
\hline
\addlinespace
 $\Gamma$$^*$  &$2.30^{+0.03}_{-0.05}$
& $2.33^{+0.05}_{-0.08}$ & $1.97^{+0.02}_{-0.02}$& $1.97^{+0.03}_{-0.03}$\\
\addlinespace
$A_{\rm cutoffpl}$ &$(1.60^{+0.06}_{-0.06})\times10^{-3}$ & $(1.56^{+0.09}_{-0.08})\times10^{-3}$ &$(3.44_{-0.08}^{+0.08})\times10^{-3}$ & $(3.47_{-0.12}^{+0.13})\times10^{-3}$\\
$E\rm _{cut}$~(keV) &   unconstrained & $<$148 & $57^{+18}_{-11}$& $38^{+11}_{-8}$\\
\addlinespace
$f_{\rm x}$ &  $0.042^{+0.012}_{-0.007}$  & $0.10^{+0.03}_{-0.03}$& $<$0.05 & 0.12$^{+0.07}_{-0.04}$\\
\addlinespace
$h_{\rm f}$& $0.13^{+0.02}_{-0.01}$ &  $<$0.19& $0.13^{+0.08}_{-0.04}$ &   $0.45^{+0.08}_{-0.11}$     \\
\addlinespace
$\tau$ &  $<$11.2    & $<$13.5 & $11.2^{+2.0}_{-0.9}$  & $>19$\\
\addlinespace
$A_{\rm reXcor}$& $(4.1^{+0.6}_{-1.0})\times10^{-16}$ &$(1.14^{+0.73}_{-0.36})\times10^{-16}$ & $(6_{-3}^{+3})\times10^{-16}$ & $(5_{-2}^{+3})\times10^{-17}$\\
\addlinespace
 \hline
$\chi^2/\mathrm{dof}$ & 1828/1725  & 1738/1725 & 1023/930 & 988/930 \\
\hline
\end{tabular}
\caption{Best-fit spectral parameters for the 2013 September and 2018 \textit{XMM-Newton} + \textit{NuSTAR} observations, modeled using two \textsc{reXcor} table models with coronal heights of 5~$R_{\mathrm{g}}$ and 20~$R_{\mathrm{g}}$.\\ (*)- \textsc{cutoffpl} and \textsc{reXcor} photon indices were tied during fitting.}
    \label{tab:XMM+NuSTAR_parametrs}
\end{table*}

\subsection{Spectral fits to the AstroSat X-ray data}

We present a detailed \textsc{reXcor} spectral analysis of the 2021 April \textit{AstroSat} observation (As3) of PDS~456 during its flaring phase. With an exposure $\sim$120~ks, As3 provides longer coverage than the 2021 March observation (As2). The 2019 observation (As1) is excluded from the analysis because of its significantly lower count rate.

\begin{figure*}[htbp]
    \centering
    \includegraphics[width=1.0\linewidth]{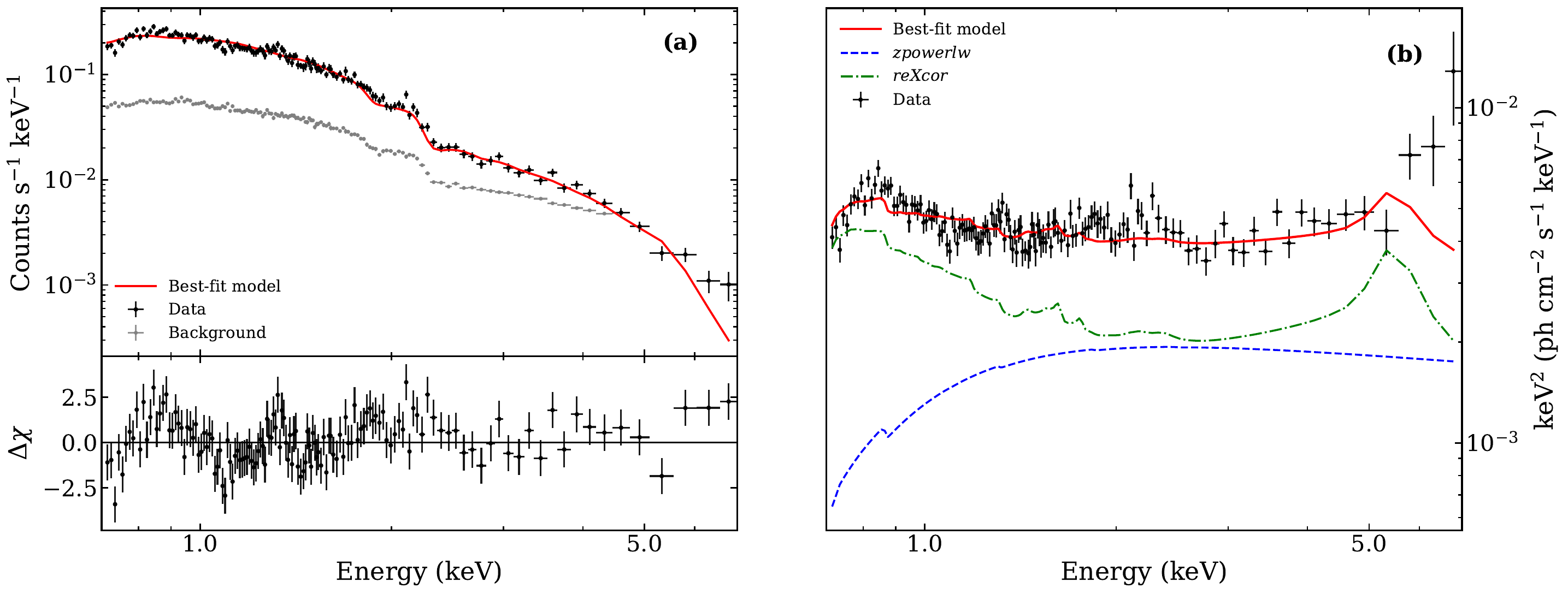}
    \caption{2021 AstroSat X-ray spectrum of PDS~456 fitted with the best-fitting \textsc{tbabs(zpowerlaw+reXcor ($l=0.1$, $a=0.99$, $h=20~R_{\rm g}$))} model. (a) The SXT spectrum and background spectrum (grey), together with the best-fitting model (red). The lower panels show the $(\mathrm{data}-\mathrm{model})/\mathrm{error}$. 
    (b) The corresponding unfolded spectrum, showing the total best-fitting model (red), \textsc{zpowerlw} (blue dashed), and \textsc{reXcor} (green dash-dotted).} 
    \label{fig: Astr_2021_plots}
\end{figure*}

The As3 spectrum is fitted over 0.7--7~keV using the same two \textsc{reXcor} models, along with zpowerlaw and tbabs. The Eddington ratio and black hole spin are fixed to the values adopted in the previous section, while the lamp-post height is explored at h = 5~$R\rm{_g}$ and h = 20~$R\rm{_g}$. The resulting best-fit spectrum for a coronal height of 20~$R\rm{_g}$ is shown in Figure~\ref{fig: Astr_2021_plots}, as a representative plot. Panel (a) shows the count spectrum, background, and residuals, while panel (b) shows the unfolded spectrum, best-fitting model, individual spectral components, and residuals. Best-fit Parameters are estimated using the same MCMC approach adopted in the previous section and are listed in Table~\ref{tab:tableAs21}. The corresponding posterior distributions from MCMC analysis are presented in Appendix \ref{App:post_dist}. For brevity, we show the posterior distributions for one low flux state (XMM13) and one high flux state (As3).

The best-fit photon index of the primary power-law is $\Gamma$ $\simeq$2.1, with the photon indices of the \textsc{zpowerlaw} and \textsc{reXcor} components tied during fitting. The optical depth $\tau$ is constrained to an upper limit of 28 at \textit{h}=~5~$R_{\rm g}$, whereas at 20~$R_{\rm g}$ only a lower limit of 19 is obtained. For {\it h}= 20~$R\rm{_g}$, the warm-coronal heating fraction dominates over the lamp-post heating fraction, whereas no clear systematic trend is observed for {\it h}= 5~$R\rm{_g}$.

\subsection{Fe K$\alpha$ versus 2-10~keV continuum}
The broadband \textsc{reXcor} analysis indicates that the disk–corona energy balance varies with flux state, with reflection more prominent at lower fluxes and the warm corona emission at higher fluxes. PDS~456 is known to exhibit a prominent broad Fe K emission feature spanning approximately 6–8~keV in the rest frame \citep{Reeves_2009, Nardini_2015, Xrism_2025}. We therefore investigate how this feature responds to changes in the underlying X-ray continuum flux.

Since the Fe K$\alpha$ emission is treated self-consistently within the above physical models, the individual line flux cannot be directly extracted. We therefore fit the 2001–2024 datasets with the phenomenological model \textsc{$M_{\rm Fe}$=tbabs(zpowerlw + zgauss)}, to examine the relation between the Fe K$\alpha$ line and 2–10 keV continuum flux. Fitting was performed over the 4–10 keV energy range to avoid contamination from the soft excess and other complex components at lower energies, while focusing on the Fe K$\alpha$ band where the primary continuum and line emission dominate. The model adequately describes all spectra, with reduced $\chi^2$ values close to unity.

\begin{figure}[htbp]
    \centering
    \includegraphics[width=0.98\linewidth]{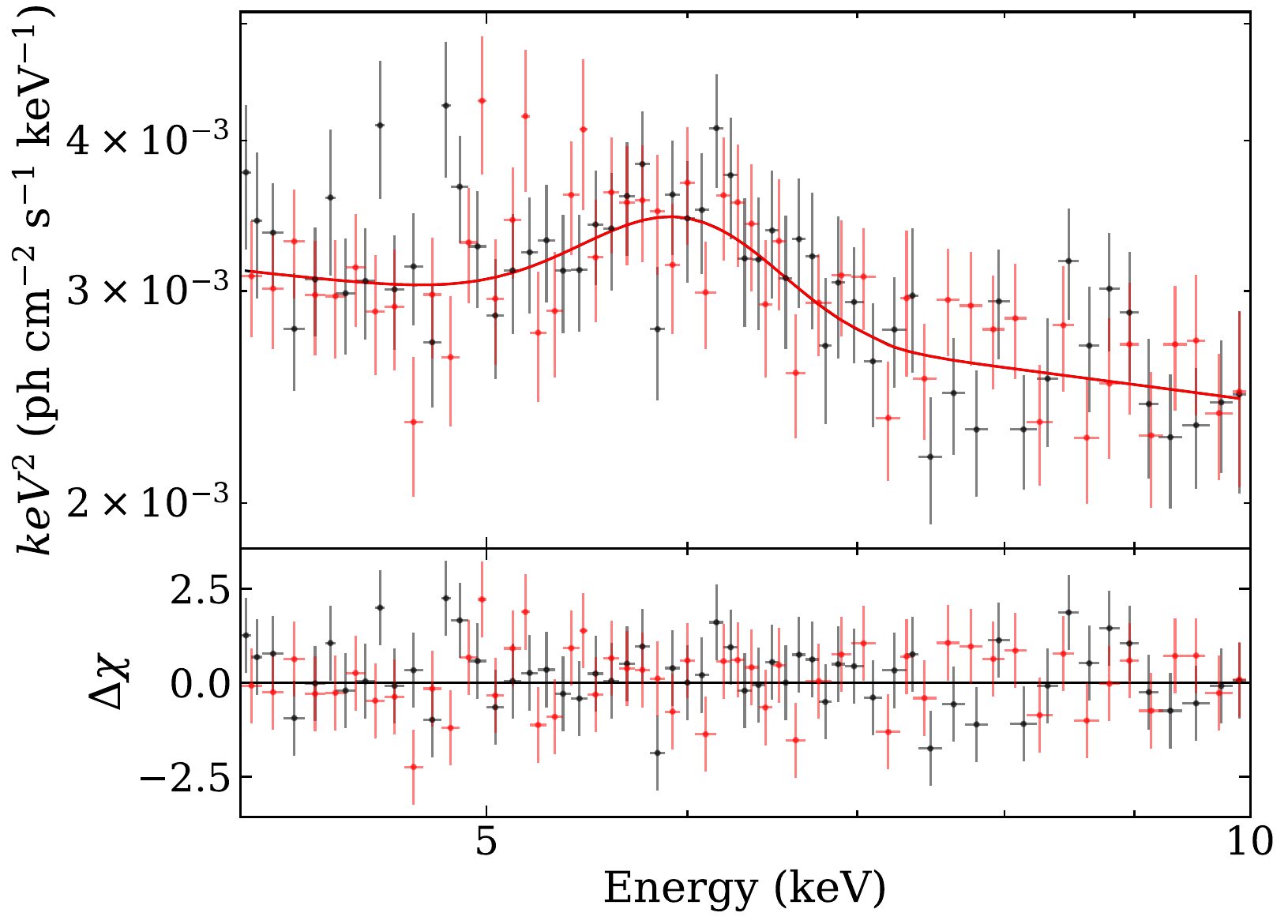}
    \caption{Broad Fe K emission line profile from the 2013 August 27 NuSTAR observation in the 4–10~keV band, using both FPMA and FPMB data shown in black and red, respectively. The spectrum is fitted with tbabs(zpo+zgauss), with the broad Fe K feature modeled using a Gaussian emission line with a best-fitting centroid energy of 6.95~keV and width $\sigma$ = 0.62~keV.}
    \label{fig:FeLine_spectra}
\end{figure}

\begin{figure}[htbp]
    \centering
    \includegraphics[width=1.0\linewidth]{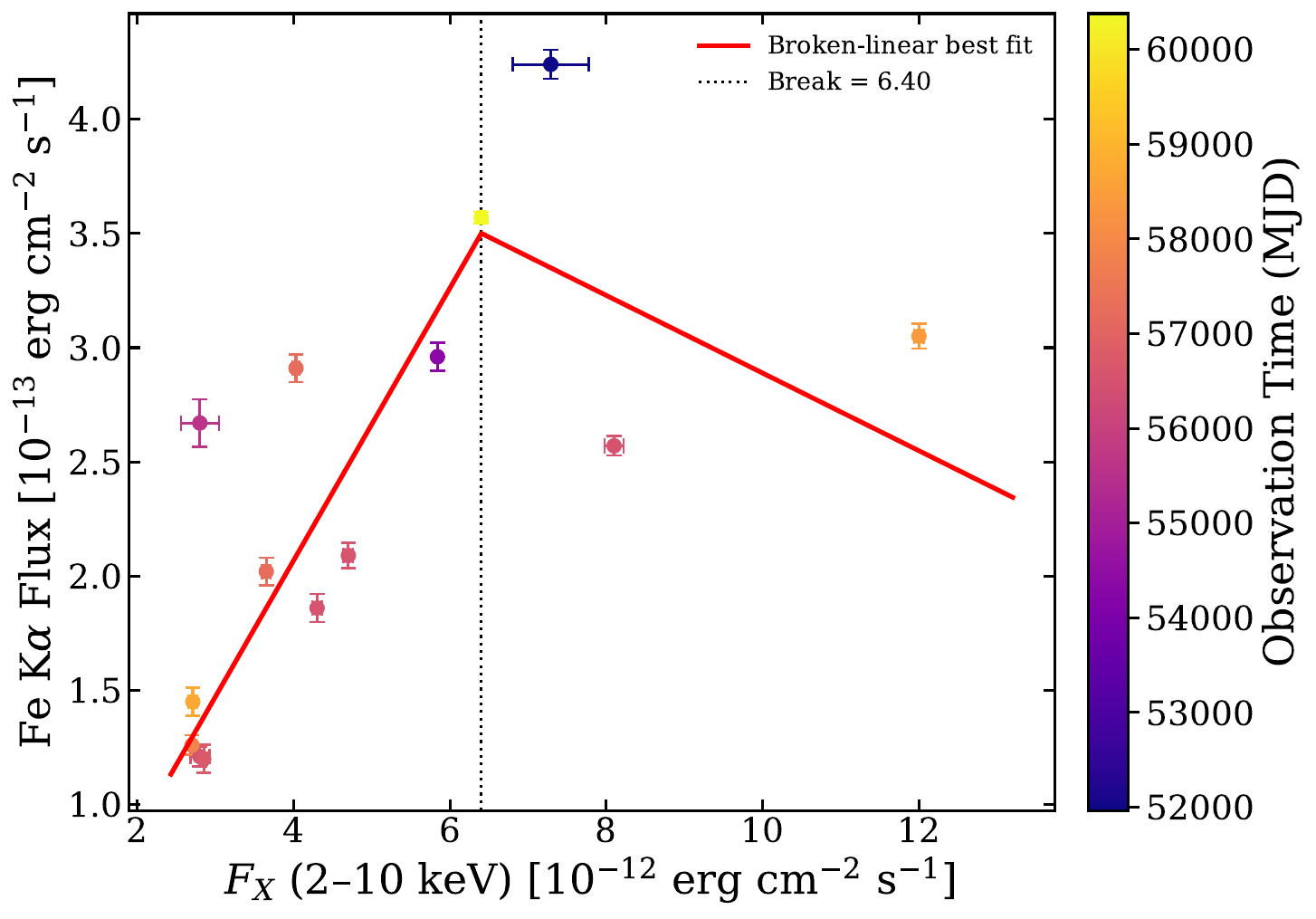}
    \caption{Flux–flux diagram showing the Fe K$\alpha$ line flux versus the 2–10~keV continuum flux for PDS~456 over the period 2001–2024. The Fe K$\alpha$ line flux was measured from the Gaussian emission component, while the continuum flux was obtained from the power-law component. The data points are color-coded according to the observation epoch (MJD), illustrating the time evolution of the source. The solid line represents the best-fit broken-linear relation, illustrating the correlation between the Fe K$\alpha$ emission and the primary X-ray continuum. The correlation weakens above a break flux of $F_{\rm cont}\sim 6.4\times10^{-12}$ erg cm$^{-2}$ s$^{-1}$.
}
    \label{fig:Fe_flux}
\end{figure}

A broad Fe K$\alpha$ emission line centered at $\sim$7~keV, with a characteristic width of $\sigma\sim$1~keV, is detected in all observations except the \textit{AstroSat} epoch, where no distinct line feature is identified. Figure ~\ref{fig:FeLine_spectra} shows a representative best-fit Fe K$\alpha$ line profile for the 2013 Nu1, consistent with previously reported broad line emission in PDS~456 \citep{Reeves_2009, Xrism_2025}. The line flux was measured using the convolution model \textsc{cflux}, applied to the Gaussian component, listed in Table~\ref{tab:Fe_parametrs}, while the 2-–10~keV continuum flux was taken from Table~\ref{tab:summary_parm}. The values were used to construct a flux–flux diagram, shown in Figure ~\ref{fig:Fe_flux}, with data points color-coded by observation epoch (MJD) to trace the source evolution across flux states. The Fe K$\alpha$ equivalent width shows no significant correlation with the 2--10~keV continuum flux.

\section{Results}
\label{sec:discussion}
We have studied the X-ray spectra of PDS~456 using \textit{AstroSat, XMM–Newton, Suzaku}, and \textit{NuSTAR} observations. Phenomenological models were first used to characterize the spectral shape and identify key spectral trends, including correlations among the primary continuum parameters. We then applied the physically motivated models to explore the broadband spectral features and the disk-corona energy balance across different flux states. 
\begin{table}[htbp]
    \centering
    \begin{tabular}{cccc}
    \hline
     Parameter  & h = 5~$R\rm{_g}$ & h = 20~$R\rm{_g}$\\
    \hline
    $\Gamma$$^*$ & $2.13_{-0.09}^{+0.07}$ & $2.10^{+0.08}_{-0.07}$ \\
    \addlinespace
    $A_{\rm zpowerlw}$ & $4_{-1}^{+1} \times 10^{-3}$ &$5.9_{-1.3}^{+0.7} \times 10^{-3}$ \\
    \addlinespace
    $f\rm{_x}$  & $0.13_{-0.02}^{+0.01}$   &  $<$0.16\\
    \addlinespace
    $h\rm{_f}$   & $<0.5$ & $0.59^{+0.06}_{-0.10}$ \\
    \addlinespace
    $\tau$    & $<$28 & $>$19 \\
    \addlinespace
    $A_{\rm reXcor}$  & $(9_{-5}^{+8}) \times 10^{-16}$ & $(5_{-2}^{+3}) \times 10^{-16}$  &\\
    \addlinespace
    \hline
    $\chi^2$/dof  & 424.84/351 & 454.52/351\\
    \hline
    \end{tabular}
    \caption{Best-fit spectral parameters for the 2021 \textit{AstroSat} observation during outburst, modeled using two \textsc{reXcor} table models with coronal heights of $5R_{\mathrm{g}}$ and $20R_{\mathrm{g}}$.\\ (*)-Powerlaw and {\sc reXcor} photon indices were tied.}
    \label{tab:tableAs21}
\end{table}
\subsection{Properties of the hot corona}
Using the $M\rm{_{1,S}}$ model, we find that PDS 456 re-brightened in 2021 following the 2018 flare \citep{Reeves_2021} and a quiescent phase in 2019. The observed flux increased from $(2.4\pm 0.3) \times 10^{-12} \mathrm{erg\,cm^{-2}\,s^{-1}}$ in 2019 to  $(5.49^{+0.12}_{-0.11}) \times 10^{-11} \, \mathrm{erg\,cm^{-2}\,s^{-1}}$ in 2021 (Figure~\ref{fig:specral_transition}), accompanied by spectral hardening. This harder-when-brighter trend, previously reported for PDS~456 \citep{Reeves-Turner_2000, Reeves_2002, Reeves_2021}, is in contrast with the softer-when-brighter behavior commonly observed in radio-quiet AGN \citep{Serafinelli_2017}. Consistently, this trend is also evident in our 2019–2021 \textit{AstroSat} analysis, where the photon index decreases from $\Gamma$ $\sim$3.0 to $\Gamma$ $\sim$2.5 as the flux increases.

To further investigate this behavior, we analyzed archival \textit{XMM-Newton, Suzaku}, and \textit{NuSTAR} data spanning 2001-2024 using the same modeling approach. The best-fit parameters (Table~\ref{tab:summary_parm}) and their temporal evolution (Figure~\ref{fig:residuals}) show that both the 2018 and 2021 outbursts coincide with comparatively lower photon indices, indicating harder spectra at higher fluxes. Moreover, the $\Gamma$– $\lambda_{\rm Edd}$  relation reinforces this softer-when-fainter behaviour. Such a trend has been observed in low-luminosity AGN (LLAGN) \citep{Gu_2009, Younes_2011, She_2018}, which are often associated with radiatively inefficient \citep{Soria_2006, Ho_2009}, hot accretion flows and truncated or absent inner thin disks, as in advection-dominated accretion flows \citep[ADAF;][]{Narayan-Yi_1995, Esin_1997, Nemme_2014}. Despite its high accretion rate \citep{Reeves_2009, Reeves_2021, Gravity_2023}, PDS~456 exhibits similar spectral behavior.
\begin{table}[htbp]
\centering
\begin{tabular}{cccccccccccc}
\hline
  MJD& Line Energy & $\sigma$ & Fe K$\alpha$ line flux \\
   & keV& keV & ($10^{-13}$~erg~cm$^{-2}$~s$^{-1}$)\\
  
  \hline
   51965.0 &  6.88$\pm$0.2 & 1.00$\pm$0.4 &  $4.24 \pm 0.11$ \\
  54355.1   & $7.01 \pm 0.21$ & $0.88 \pm 0.25$ & 2.96$\pm$0.1\\
  55637.0  & $6.84\pm 0.22$ & $1.2 \pm 0.92$ & 2.67$\pm$0.17\\
  56531.2  & $6.95 \pm 0.15$ & $0.62 \pm 0.17$ & 2.57$\pm$0.07\\
  56541.1 & $7.07 \pm 0.18$ & $0.51 \pm 0.21$& 1.21$\pm$0.07\\
  56550.7 & $6.97 \pm 0.17$ & $0.62 \pm 0.18$& 1.86$\pm$0.1\\
  56555.1 & $6.94 \pm 0.13$ & $0.68 \pm 0.15$ & 2.09$\pm$0.09\\
  56714.3 & $6.6 \pm 0.23$ & $0.76 \pm 0.23$ & 1.20$\pm$0.1\\
  57224.5 & $6.37 \pm 0.68$ & $1.12 \pm 0.53$ & 2.91$\pm$0.1 \\
  57227.5 & $6.99 \pm 0.25$ & $0.87 \pm 0.26$ & 2.02$\pm$0.1\\
  57835.2 & $6.86 \pm 0.09$ & $0.62 \pm 0.09$ & 1.26$\pm$0.07\\
  58381.1 & $6.93 \pm 0.18$ & $0.82 \pm 0.19$ & 3.05$\pm$0.09 \\
  58750.0 & $6.78 \pm 0.19$ & $0.77 \pm 0.19$ & 1.45 $\pm$ 0.09\\
  60379.9& $6.88 \pm 0.06$ & $0.72 \pm 0.07$ & 3.57 $\pm$ 0.04 \\ 
  \hline
\end{tabular}
    \caption{Summary of the Fe K$\alpha$ emission-line parameters obtained from phenomenological spectral modeling using \textsc{tbabs(zpowerlw+zgauss)} in the 4–10 keV energy band. The line energy and width are given with their corresponding uncertainties, and Fe K$\alpha$ fluxes are also listed. Fe K$\alpha$ parameters are not reported for the \textit{AstroSat} observations, as no significant Fe K$\alpha$ feature could be identified.
}
    \label{tab:Fe_parametrs}
\end{table}

 This contrast makes the source particularly intriguing and raises important questions about the nature of its accretion flow in the high-accretion regime. This suggests that analogous accretion flow conditions or coronal properties can arise even in high accretion rate systems, where the inner flow may become optically thick, trapping a significant fraction of the radiation and advecting it inward, often described as a slim disk \citep{Abramowicz_1988, Narayan_1998}. Such advection-dominated inner regions are also capable of driving powerful outflows \citep{Nemmen_2007, Begelman_2012}. 

The observed softer-when-fainter trend in PDS~456 may indicate the presence of an efficiently cooling corona during low flux states. This interpretation is consistent with the coronal temperature inferred by \cite{Reeves_2021}, who reported a high-energy cutoff of $\sim$50~keV from the 2018 \textit{XMM–Newton} and \textit{NuSTAR} observations, corresponding to a relatively low coronal temperature of $\sim$15~keV, considerably lower than the typical values observed in most AGN. The prominent big blue bump \citep{Simpson_1999} suggests a substantial supply of soft seed photons, which could efficiently cool the corona and naturally account for the observed spectral steepening at lower flux levels. 

Building on our results, we employ the physically motivated \textsc{reXcor} model to investigate the broadband X-ray spectra of PDS~456 across multiple epochs. The model provides a self-consistent treatment of ionized relativistic reflection and warm coronal emission, accounting for relativistic effects and the distribution of accretion power between the compact lamp-post X-ray source and the warm corona. Our analysis of the 2013, 2018, and 2021 observations reveals clear variations in the coronal energy distribution.

In 2013, corresponding to a low-flux state, the lamp-post heating fraction increases with coronal height, while the warm corona contribution decreases, indicating that the compact lamp-post corona dominates the energy deposition during faint states. In contrast, the 2018 high-flux observations show that the warm corona becomes energetically dominant at larger coronal heights, reflecting a shift toward extended coronal heating during flaring states. The 2021 {\it AstroSat} observation further reinforces this trend, with the warm corona dominating the energy output for a coronal height of 20~$R\rm{_g}$. At a coronal height of 5~$R\rm{_g}$, no systematic variation is observed across the three epochs. The photon index ($\Gamma$) from the {\sc reXcor} modeling also shows a corresponding evolution, being steeper in 2013 ($\sim$2.3) and harder in 2018 and 2021 observations ($\sim$1.9 and $\sim$2.1). This behavior is consistent with the softer-when-fainter trend and suggests that variations in coronal geometry and the distribution of accretion power directly influence the spectral shape. A qualitatively similar flux-dependent coronal evolution has been reported in the NLS1 Mrk~335, with a compact, reflection-dominated corona in the low-flux state and a more continuum-dominated one in the high-flux state \citep{1Gallo_2014}. Although not framed in terms of warm-corona and lamp-post energy partition, this behavior is broadly consistent with our results.

\subsection{Fe K$\alpha$ response}
The spectral modeling presented above establishes that PDS~456 undergoes a transition from reflection-dominated states at low flux to warm-corona-dominated states at high flux, while remaining in a high-accretion slim disk regime. An independent test of this evolving disk–corona energy balance can be obtained from the variability of the Fe K$\alpha$ emission line, which directly traces the response of the reflecting disk to the illuminating continuum.

The physical origin of the broad line feature, however, may involve both relativistic reflection from the inner accretion disk and reprocessing in the disk wind. In particular, broad Fe K emission in PDS 456 has been interpreted in terms of wind-related Compton scattering \citep{Matzeu_2022}, while reflection-based models can also provide a good description of the observed Fe K profile \citep{Parker_2022}. \cite{Parker_2022} further demonstrated that disk-reflection and disk-wind models can be highly degenerate when constrained by Fe K spectroscopy alone, making their relative contributions difficult to disentangle. We therefore use the phenomenologically measured Fe K$\alpha$ line flux to investigate its response to variations in the 2–10~keV continuum flux.

We find a clear positive correlation between the Fe K$\alpha$ line flux and the 2–10~keV continuum flux over low to intermediate flux states, indicating that the Fe K emission responds to variations in the illuminating continuum. This behavior is consistent with a contribution from disk reflection, in which hard X-ray photons from the corona irradiate the accretion disk and produce fluorescent emission \citep{George_1991, Reynolds_2003}. The strong correspondence between line and continuum flux in the reflection-dominated low-flux states is consistent with efficient disk illumination by a compact corona. However, given the degeneracy between disk reflection and wind reprocessing, a contribution from the disk wind to the observed Fe K emission cannot be ruled out.

At the highest flux levels, the Fe K$\alpha$ line flux shows a weaker response to the continuum. To quantify this trend, we compared linear and broken-linear models. The broken-linear model is preferred based on comparative fit statistics and indicates that the line flux increases with continuum flux up to a break at $F_{\rm cont}\sim 6.4\times10^{-12}$ erg cm$^{-2}$ s$^{-1}$, followed by a declining trend at higher continuum fluxes. This behavior coincides with the transition to warm-corona-dominated spectra as per the \textsc{reXcor} modeling. In this regime, a larger fraction of the accretion power is dissipated within the corona itself rather than being reprocessed by the disk, leading to enhanced intrinsic continuum emission but a relatively weaker reflected component. This change in the disk–corona energy balance explains the reduced responsivity of the Fe K$\alpha$ emission at high flux levels and supports a decreasing contribution of reflection as the warm corona becomes energetically dominant.

The color-coding of the data points by observation epoch indicates that deviations from the linear relation are not time-dependent but are instead linked to the flux state of the source. Low flux observations from different epochs remain consistent with the best-fit, showing that the Fe K$\alpha$ line continues to track the continuum whenever the source is in a low flux state. This suggests that the weakening of the correlation at high flux levels is driven by the changes in the flux state rather than long-term temporal evolution.

\section{Conclusions}
\label{sec:conclusion}

We have performed a broadband X-ray spectral study of PDS~456, revealing significant evolution in its coronal properties and disk–corona energy balance across multiple flux states over 2001–2024. We find strong long-term spectral variability, including a clear softer-when-fainter trend across the full dataset. The increase in photon index during low-flux states suggests more efficient coronal cooling when the source is faint, possibly driven by the strong soft photon field associated with its prominent big blue bump. The physically motivated {\sc reXcor} modeling shows that the distribution of accretion power between the compact lamp-post corona and the warm Comptonizing region evolves with flux. At low flux, the spectrum is reflection-dominated, indicating efficient disk illumination by a compact corona, while at high flux, the warm corona becomes energetically dominant, implying that a larger fraction of the accretion power is dissipated within the coronal structure rather than being reprocessed by the disk. We also find a clear positive correlation between the Fe K$\alpha$ emission line flux and the 2-10~keV continuum at low to intermediate flux levels, indicating that the Fe line responds to the illuminating continuum. At the highest flux levels, this correlation weakens, suggesting a change in the response of the Fe line as the contribution from the warm corona becomes enhanced. This indicates that the Fe K$\alpha$ variability is primarily governed by flux state rather than long-term temporal evolution.
\begin{acknowledgments}
The authors thank the anonymous referee whose comments and suggestions have significantly improved the manuscript. FF acknowledges Dr. Jeena K. for her initial guidance in {\it AstroSat} data reduction and for providing access to her laboratory facilities.
RC and AS thank ISRO for support under the \textit{AstroSat} archival data utilization program, acknowledge financial support from ANRF through a SURE grant (SUR/2022/001503), and IUCAA for their hospitality and usage of their facilities during their stay at different times as part of the university associateship program.
RC thanks Presidency University for support under the Faculty Research and Professional Development (FRPDF) Grant, and acknowledges financial support from ANRF through an ARG grant (ANRF/ARG/2025/003315/PS).
This study made use of data from the {\it AstroSat} mission, conducted by the Indian Space Research Organisation (ISRO) and archived at the Indian Space Science Data Centre (ISSDC). The analysis also utilized data from the Soft X-ray Telescope (SXT), developed at TIFR, Mumbai.
The SXT POC at TIFR
was instrumental in verifying and releasing the data through the ISSDC archive.
This work is partially based on observations obtained with {\it XMM-Newton}, an ESA science mission with instruments and contributions directly funded by ESA Member States and NASA.
This research has made use of data from the {\it NuSTAR} mission, a project led by the California Institute of Technology, managed by the Jet Propulsion Laboratory, and funded by the National Aeronautics and Space Administration. Data analysis was performed using the {\it NuSTAR} Data Analysis Software (NuSTARDAS), jointly developed by the ASI Science Data Center (SSDC, Italy) and the California Institute of Technology (USA).
This research has made use of data and/or software provided by the High Energy Astrophysics Science Archive Research Center (HEASARC), which is a service of the Astrophysics Science Division at NASA/GSFC.
This research has made use of the NASA/IPAC Extragalactic Database (NED), which is operated by the Jet Propulsion Laboratory, California Institute of Technology, under contract with the National Aeronautics and Space Administration.

\end{acknowledgments}
\section{Data availability} \label{sec:data_avl}
The data analyzed in this work are publicly available in the {\it AstroSat, XMM-Newton, \rm{and} NuSTAR} archives.
\facilities{AstroSat(SXT), XMM-Newton, and NuSTAR}

\software{HEASoft\footnote{\url{https://heasarc.gsfc.nasa.gov/docs/software/heasoft/}}, 
XSPEC\footnote{\url{https://heasarc.gsfc.nasa.gov/xanadu/xspec/}} \citep{Arnaud_1996, Dorman_2001, Dorman_2003}}

\appendix

\section{\textsc{reXcor} posterior distributions}
\label{App:post_dist}
\begin{figure*}
    \centering
    \includegraphics[width=\linewidth]{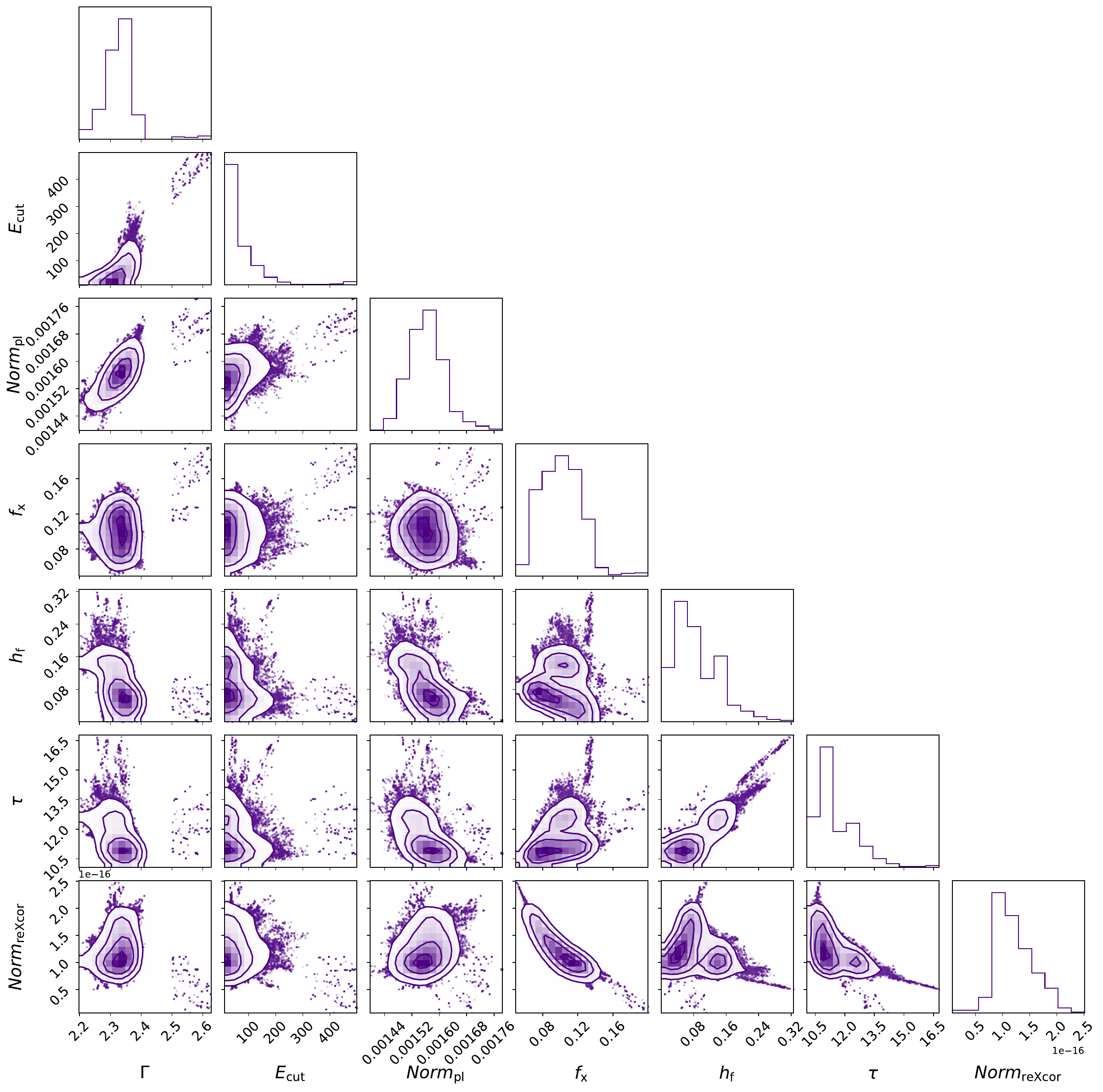}
    \caption{Corner plot showing the marginalized posterior probability distributions and pairwise correlations of the spectral model parameters inferred from MCMC samples using the {\sc reXcor} table with h = 20~$R_{\rm g}$ for 2013 joint {\it XMM-Newton} + {\it NuSTAR} observation. Diagonal panels show the 1-D posterior distributions for each parameter, while the off-diagonal panels display the corresponding 2-D joint distributions with credible interval contours, illustrating parameter degeneracies and uncertainties. During the low-flux state in 2013 the warm corona fraction, $h_{\rm f}$ is relatively lower.
    }
    \label{fig:xmm2013_h20_corner}
\end{figure*}

\begin{figure*}
    \centering
    \includegraphics[width=\linewidth]{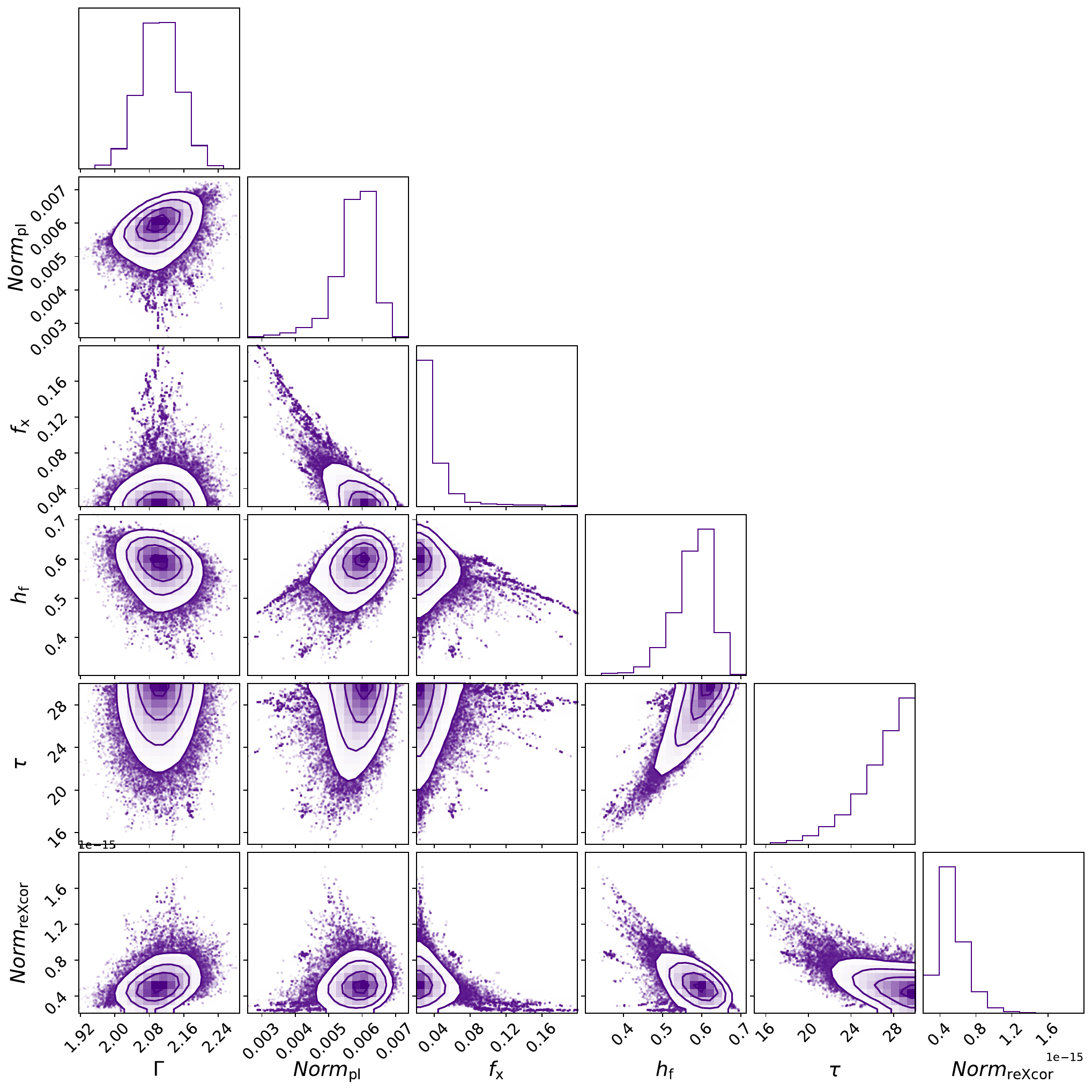}
    \caption{Same as Figure \ref{fig:xmm2013_h20_corner} using {\sc reXcor} table with h = 20~$R_{\rm g}$ for 2021 {\it AstroSat} observation. During the high-flux state in 2021 the warm corona fraction, $h_{\rm f}$ is relatively higher.}
    \label{fig:ast2021_corner}
\end{figure*}



\bibliography{APJ-PDS456}{}
\bibliographystyle{aasjournalv7}
\end{document}